\documentclass[10pt,twocolumn,journal]{IEEEtran}
\usepackage{amsmath,amssymb,amsfonts}
\usepackage{algorithmic,algorithm}
\usepackage{graphicx}
\usepackage{textcomp}
\usepackage{xcolor}
\usepackage{hyperref}
\usepackage{editing}
\begin{document}

\title{Study of Iterative Detection and Decoding for RIS-Aided Multiuser Multi-Antenna Systems

}
\author{{Roberto C. G. Porto}
{\textit{Centre for Telecommunications Studies} \\
\textit{Pontifical Catholic University of Rio de Janeiro}\\
Rio de Janeiro, Brazil \\
camara@aluno.puc-rio.br}
\and
{Rodrigo C. de Lamare}
{\textit{Centre for Telecommunications Studies} \\
\textit{Pontifical Catholic University of Rio de Janeiro}\\
Rio de Janeiro, Brazil \\
delamare@puc-rio.br} 
}
\maketitle

\begin{abstract}
We present a novel iterative detection and decoding (IDD) scheme for Reconfigurable Intelligent Surface (RIS)-assisted multiuser multiple-antenna systems. The proposed approach introduces a joint iterative detection strategy that integrates Low-Density Parity-Check (LDPC) codes, RIS processing and iterative detection and decoding. In particular, we employ a  minimum mean square error receive filter that performs truncation at the RIS and soft interference cancellation at the receiver. Simulation results evaluate the system's overall capacity and bit error rate, and demonstrate substantial improvements in bit error rate across block-fading channels.
\end{abstract}


\section{Introduction}

In the ever-evolving landscape of wireless communication systems, the escalating demand for higher data rates, improved coverage, and reduced power consumption persists. Addressing these challenges requires exploring cutting-edge technologies such as large-scale multiple-antenna systems \cite{mmimo,wence} and the concept of Reconfigurable Intelligent Surfaces (RIS). These reflective surfaces exhibit significant potential for optimizing wireless networks and are expected to play a crucial role in the development of the sixth generation of communication systems (6G) \cite{rista}.
The fundamental concept of RIS-assisted systems diverges from the conventional approach of altering the transmitted signal to reshape the wireless environment's propagation properties. Instead, it focuses on modifying the characteristics of the equivalent channel between users and the access point \cite{9140329}. The adoption of RIS holds the potential to extend coverage \cite{9359653}, reduce power consumption \cite{9548940}, and increase channel capacity \cite{9998527,9913356}.

Several works propose modifications to the traditional RIS system to address the practical issues associated with passive RIS. In \cite{9998527}, an active RIS is proposed, with the capability not only to change phase but also to amplify reflected signals. In \cite{9913356}, the authors introduce a reflective surface whose elements are fully- and group-interconnected. This design not only allows for characterization by scattering parameters but also enhances the performance of the system by enabling a beyond-diagonal reconfigurable intelligent surface.

In \cite{9452133}, the authors have studied the performance of an RIS-assisted wireless communication system where the signal from the transmitter is protected by LDPC codes. Instead of addressing the performance of the system by maximizing the Signal-to-Noise-and-Interference Ratio (SNIR) at the receiver \cite{9548940,9998527,9913356,9110912}, the most common form of optimization, this work uses bit-error rate (BER) as a benchmark for performance evaluation, contrasting Monte Carlo simulation results with analytic bounds. 
Meanwhile, in \cite{1494998, 8240730, 7105928}, the authors evaluate the iterative detection and decoding (IDD) algorithm using LDPC codes for multiple-antenna systems under block-fading channels.

In this work, we investigate the uplink of multiuser multiple-antenna systems that employ RIS, channel coding and iterative processing techniques \cite{idd_ris1,idd_ris2}. In particular, we present a novel approach by formulating a problem that aims to minimize the mean squared error between a transmitted symbol and an estimated symbol at the receiver while incorporating an IDD scheme \cite{1494998}.
By employing an IDD scheme, the detector and decoder exchange soft information, commonly represented in log likelihood ratios (LLRs). Previous research on MIMO and IDD schemes has explored block-fading channels \cite{8240730, 7105928} and LDPC codes. This information exchange results in performance enhancements, with the degree of improvement growing proportionally to the number of information exchanges. We evaluate the performance of a RIS-aided multiuser multiple-input single-output (MU-MISO) system in typical wireless scenarios based on the 3GPP standard \cite{access2010further}. By resorting to constraint relaxation and alternating optimization, we propose a joint iterative detection approach that combines LDPC decoding and RIS. Numerical results evaluate the proposed system in terms of BER and sum-rates.

The rest of this paper is organized as follows. In Section II we describe the system model. Section III describes the derivation of the proposed RIS-MMSE detector and decoding technique. Section IV discusses the simulation results and Section V concludes this paper.

\textit{Notations:} Bold capital letters indicate matrices while vectors
are in bold lowercase. $\mathbf{I_n}$ denotes a $n \text{x} n$ identity matrix and diag($\mathbf{A}$) is a diagonal matrix only containing the diagonal elements of $\mathbf{A}$. $\mathbb{C}$ and $\mathbb{R}$ denote the sets of complex and real numbers, respectively; [·]$^{-1}$, [·]$^{T}$ and
[·]$^H$ denote the inverse, transpose, and conjugate transpose operations, respectively; 

\section{System Model}

\begin{figure}[htbp]
    \centerline{\includegraphics[width=0.35\textwidth]{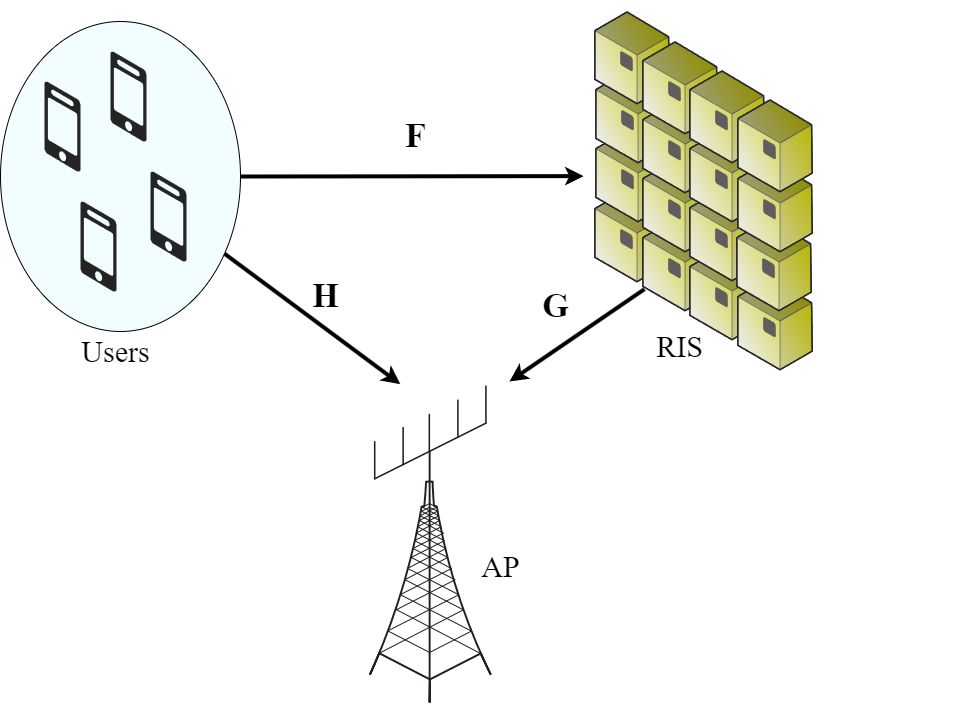}}
    \caption{An RIS-aided MISO uplink communication system.}
    \label{fig01}
\end{figure}

Consider a scenario in a single-cell multi-user uplink system featuring multiple antennas \cite{mmimo} and assisted by a RIS, as illustrated in Fig. \ref{fig01}. In this setup, there are $K$ single-antenna users, and the receiver is equipped with $M$ antennas. Each user's information symbols are firstly encoded by each user’s channel encoder and modulated to $x_k$ according to a given modulation scheme. The transmit symbols $x_k$ have zero mean and share the same energy, with $E[|x_k|^2] = \sigma^2_x$. Subsequently, these modulated symbols are transmitted over block-fading channels.

We assume that the RIS incorporates $N$ reflecting elements, with matrices $\mathbf{H}$ and $\mathbf{F}$ representing the links for MU-AP and MU-RIS, respectively. The links between the RIS and AP are characterized by the matrix $\mathbf{G} \in \mathbb{C}^{M\text{x}N}$. The RIS has $N$ reflecting units, and their reflection coefficients are modeled as a complex vector $\boldsymbol{\varphi} \triangleq [e^{j\theta_1}, \dots, e^{j\theta_N}]^T$, where $\theta_n$ represents the phase shift of the $n$th unit for all $n \in \mathcal{N} \triangleq \{ 1, 2, \dots, N\}$.
The signal model of an $N$-element passive RIS widely used in the literature is given by a diagonal phase shift matrix $\Phi \in \mathbb{C}^{N\text{x}N}$ defined as $\mathbf{\Phi} \triangleq \text{Diag}(\boldsymbol{\varphi})$.
\vspace{-1em}
%\vspace{-0.5em}

The effective channel, denoted as $\mathbf{H_{eff}} \in \mathbb{C}^{M\text{x}K}$, between the users and the AP is given by 
\begin{equation}
    \mathbf{H_{eff}} = \mathbf{H} + \mathbf{G}\mathbf{\Phi}\mathbf{F}.
    \label{Heff}
\end{equation}

The signal received $\mathbf{y}$ at AP can be represented as:
\begin{equation}
    \mathbf{y} = \mathbf{H_{eff}}\mathbf{x} + \mathbf{n},
\end{equation}
where the vector $\mathbf{x} = \{x_1, \dots, x_k\}$ represent the signal transmitted for each user and the vector $\mathbf{n} \sim \mathcal{CN}(\mathbf{0_M},\sigma_n^2\mathbf{I_M})$ represents the noise.

An estimate $\hat{x}_k$ of the transmitted symbol on the $k$th user without soft interference cancellation (SIC) is obtained by applying a linear receive filter $\mathbf{w_k}$ to $\mathbf{y}$:

\begin{equation}
    \hat{x}_k = \mathbf{w_k}^H\mathbf{y}
    \label{detection_estimate_1}
\end{equation}

\subsection{Enhancing Detection Estimates through SIC}

\begin{figure*}
\vspace{-2em}
    \centerline{\includegraphics[width=1\textwidth]{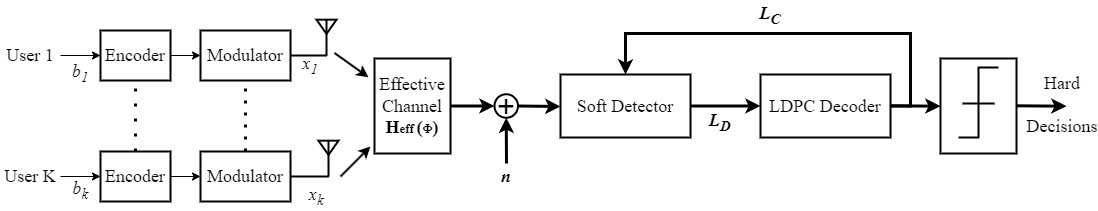}}
    \caption{System model of an IDD multiuser multiple-antenna system.}
    \label{fig02}
\end{figure*}

An IDD scheme with a soft detector and LDPC decoding is used to enhance the performance of the system. The soft detector incorporates extrinsic information provided by the LDPC decoder ($\mathbf{L_C}$), and the LDPC decoder incorporates soft information provided by the MIMO detector ($\mathbf{L_D}$), as illustrated in Fig. \ref{fig02}.

Let us denote $\mathbf{\tilde{x}}=[\tilde{x}_1,\dots,\tilde{x}_K]^T$ and
\begin{equation} 
    \tilde {\mathbf {x}}_{k} = \tilde {\mathbf {x}} - \tilde {x}_{k}\mathbf {e}_{k},
\end{equation}
where $e_k$ is a column vector with all zeros, except that the $k$th element is equal to 1. For each user $k$, the interference from the other $K$ - 1 users is canceled according to
\begin{equation} 
    \mathbf {y}_{k}=\mathbf {y}-\sum _{j=1,j\neq k}^{K}\tilde {x}_{j}\mathbf {{h}_{j}^{eff}}=\mathbf {y}-\mathbf {H_{eff}}\tilde {\mathbf {x}}_{k}. 
\end{equation}
where $\mathbf {h}_{j}$ represents the lines of the matrix $\mathbf{H_{eff}} = [\mathbf{{h}_{1}^{eff}}, \dots, \mathbf{{h}_{K}^{eff}}]^T$.

A detection estimate $\hat{x}_k^{\rm sic}$ of the transmitted symbol on the $k$th user is obtained by applying a linear filter $\mathbf{w_k}$ to $\mathbf{y_k}$:
\begin{equation}
    \hat{x}_k^{\rm sic} = \mathbf{w_k}^H\mathbf{y_k}
    \label{sic}
\end{equation}

\subsection{Sum-Rate for RIS-MISO systems}

The estimation of the Sum-Rate without the SIC relies on the SINR computed for each user using the MU-MISO transmission model presented in (\ref{detection_estimate_1}). The SINR for the $k$-th user, denoted as $\gamma_k$, is given by

\begin{equation} 
    {\gamma _{k}} = \frac {{{{\left |{ {{{ {\mathbf w}}^{\mathrm{ H}}_{k}}{{\mathbf{h}}_{k}^\text{eff}}} }\right |}^{2}}}\sigma_x ^{2}}{{\sum \nolimits _{j = 1,j \ne k}^{K} {{{\left |{ {{{ {\mathbf w}}^{\mathrm{ H}}_{k}}{{\mathbf{h}}_{j}^\text{eff}}} }\right |}^{2}}\sigma_x ^{2}} +  ||{\mathbf w}^{\mathrm{ H}}_{k}||^2 \sigma_n ^{2}}}. 
\end{equation}

When considering the scenario with SIC as expressed in (\ref{sic}), the output of the SIC-MMSE filter can be approximated as a complex Gaussian distribution due to the large number of independent variables \cite{8240730} by:

\begin{equation}
    \hat{x}_k^{\rm sic} = \mu_kx_k + z_k
    \label{eq:approx}
\end{equation}
where $\mu_k$ is the equivalent amplitude of the $k$ user's signal, and the parameter $z_k$ represents the combination of noise and the residual intersymbol interference, which is equivalent to $z_k \sim \mathcal{N}(0,\eta_k^2)$. Using this equation, the SINR at the output of the soft instantaneous MMSE filter can be defined as \cite{774855}:

\begin{equation}
     {\gamma _{k}^\text{sic}} \triangleq   
     \frac{E[(\hat{x}_{k}^{\text{sic}})^2]}{\text{var}[\hat{x}_{k}^{\text{sic}}]} = 
     \frac{\mu_k^2\sigma_x^2}{\eta_k^2},
\end{equation}
and the corresponding Sum-Rate, denoted as $R_{\mathrm{sum}}$, is then calculated as:
\begin{equation}
    R_{\mathrm{sum}}= \sum \limits _{k = 1}^{K} {\log _{2}\left ({{1 + {\rm SINR_{k}}} }\right)}
    = \sum \limits _{k = 1}^{K} {\log _{2}\left ({{1 + {\gamma _{k}^\text{sic}}} }\right)}.
\end{equation}

\section{Joint Iterative Detection with LDPC Decoding and Design of Reflective Parameters}

The incorporation of RIS introduces challenges to the design of an uplink receiver filter, primarily due to its dependence on the diagonal phase shift matrix ($\mathbf{\Phi}$). Additionally, the adjustment of the phase shift matrix requires information from the receive filter. To address this issue, we employ a scheme based on alternating optimization, which will be detailed in the subsequent sections.

\subsection{Design of Reflection Parameters}
\label{sec:designRIS}
We employ the symbol estimate without SIC from (\ref{detection_estimate_1}). Our problem formulation begins by fixing the value of the reception filter and seeking the matrix $\mathbf{\Phi}$ that minimizes the following expression:

\begin{align*} 
\underset{\boldsymbol \Phi }{{\text{minimize}}} \quad \ &
E[||\mathbf{x} - \hat{\mathbf{x}}^2||^2_2] = E[||\mathbf{x}-\mathbf{W}( \mathbf{H_{eff}}\mathbf{x} + \mathbf{n})||^2_2]
\\ {\text{subject to}}\quad \ & \mathbf{\Phi}=\text{diag}\{e^{j\theta_1}, \dots, e^{j\theta_N}\}
\\ & 0 \leq \theta_n < 2\pi, n=1,\dots, N
\\ & \text{tr}(\mathbf{R}_x) = \sigma_x^2\textbf{I}
\end{align*}
where $\mathbf {W} \triangleq [\mathbf{w_1}, \dots, \mathbf{w_K}]^T $ represents the reception filter matrix.

While the norm consistently exhibits convex behavior, it is noteworthy that the restriction on diagonal elements of the matrix $\mathbf{\Phi}$ do not form a convex set \cite{tds1,tds2}. To address this, we employ a relaxation of this constraint. However, the solution to the problem results in a full-rank matrix $\mathbf{\Phi}$. 

For an efficient algorithmic solution, we use the reflection coefficients as a complex vector diag$(\boldsymbol{\varphi})=\mathbf{\Phi}$ and reformulate the (\ref{Heff}) in the following equivalent form
\begin{equation}
    \mathbf{h}_{k}^{\rm eff} = \mathbf{h}_{k} + \mathbf{GA_k}\boldsymbol{\varphi},
\end{equation}
where $\mathbf{A_k} = \text{diag}(\mathbf{f_k})$ and $\mathbf{f_k}$ represent the links between the user $k$ and the RIS. Additionally, we have $\mathbf{F} = [\mathbf{f}_{1}, \dots, \mathbf{f}_{K}]^T$ which leads to
\begin{equation}
    E[||\mathbf{x} - \hat{\mathbf{x}}^2||^2_2] = E[||\mathbf{x}-\mathbf{W}(\sum_{i=1}^K \mathbf{h}_{i}^{\rm eff}(\boldsymbol{\varphi})\mathbf{x}_i + \mathbf{n})||^2_2].
\end{equation}

By computing the derivative of the mean-square error (MSE) cost function with respect to $\boldsymbol{\varphi}$ and setting the result to zero yields:

\begin{equation}
\boldsymbol{\varphi}_o\mathbf{ = {\rm diag}(\boldsymbol{\beta}^{-1}.\Psi)},
\label{eq:phi}
\end{equation}
where 
\begin{equation}
    \boldsymbol{\beta} = \sum_i^K\mathbf{(WGA_i)^H(WGA_i)}    
\end{equation}
\begin{equation}
    \boldsymbol{\Psi} = \sum_i^K\mathbf{(WGA_i)^H(e_i-Wh_i^{\rm eff})}.
\end{equation}
Since we have opted for the relaxation of the constraint for MMSE processing \cite{jidf,idd_ris1,idd_ris2}, it becomes necessary to truncate the computed value by

\begin{equation}
    \boldsymbol{\varphi_t} = \frac{[{\varphi}_o]_i}{|[{\varphi}_o]_i|} = e^{j\measuredangle (\boldsymbol{\varphi_0})}.
\end{equation}

\subsection{Iterative Detection and LDPC Decoding}

Inspired by prior work on IDD schemes \cite{1494998,8240730,spa,mfsic,mbdf,dfcc,did,1bitidd,listmtc,detmtc,msgamp1,msgamp2,comp}, the soft estimate of the $k$th transmitted symbol is firstly calculated based on the $\mathbf{L_c}$ (extrinsic LLR) provided by the channel decoder from a previous stage:

\begin{equation*} \tilde {x}_{k}=\sum _{x\in \mathcal {A}}x\text {Pr}(x_{k}=x)=\sum _{x\in \mathcal {A}}x\left ({\prod _{l=1}^{M_{c}}\left [{1+\text {exp}(-x^{l}L_{c}^{l})}\right]^{-1}}\right), \end{equation*}
where $\mathcal {A}$ is the complex constellation set with $2^{M_c}$ possible points. The symbol $x^l$ corresponds to the value $(+1, -1)$ of the $l$ th bit of symbol $x$ . 

A detection estimate uses SIC from (\ref{sic}), where the value of $\mathbf{\Phi}$ is fixed and $\mathbf{w_k}$ is chosen to minimize the mean square error (MSE) between the transmitted symbol $x_k$ and the filter output
\begin{equation} 
    \mathbf {w}_{k}=\arg \min _{\mathbf {w'}_{k}} E\left [{\left \vert{ x_{k}-\mathbf {w'}_{k}^{H}\mathbf {y}_k}\right \vert ^{2}}\right]. 
\end{equation}
It can be shown that the solution is given by
\begin{equation}
    \mathbf{w_k} = \left(\frac{N_0}{E_x}\mathbf{I_{n_r}} + \mathbf{\mathbf{H_{eff}} \Delta_k \mathbf{H_{eff}}}^H \right)^{-1}\mathbf{h_k^{eff}},
    \label{eq:w}
\end{equation}
where the covariance matrix $\mathbf{\Delta_k}$  is
\begin{equation}
    \mathbf{\Delta_k} = \text{diag}\left[\frac{\sigma^2_{x_{1}}}{E_x}\dots \frac{\sigma^2_{x_{k-1}}}{E_x}, 1, \frac{\sigma^2_{x_{k+1}}}{E_x},\dots,\frac{\sigma^2_{x_{K}}}{E_x}  \right],
\end{equation}
$E_x$ is the average energy of the transmitted symbol and $\sigma^2_{x_{i}}$ is the variance of the $i$th user computed as:
\begin{equation}
\sigma_{x_{i}}^{2}=\sum\limits_{x\in {\cal A}}\vert x-\bar{x} _{i}\vert ^{2}P(x_{i}=x). 
\end{equation}

Using the approximation presented in  (\ref{eq:approx}), the mean and variance of the estimated symbol $\hat {x}_{k}^\text{C}$, conditioned on the transmitted symbol $x$, are provided as follows:
\begin{equation}
    \mu _{k}\overset {\Delta }{=} E\left [{\hat {x}_{k}^\text{sic}\vert x}\right] = \mathbf{w^H_k}\mathbf{h_k^{eff}}. 
\end{equation}
\begin{equation}
    \eta _{k}\overset {\Delta }{=} \text{var}\left [{\hat {x}_{k}^\text{sic}\vert x}\right] = E_x(\mu _{k}-\mu _{k}^2).
\end{equation}

Therefore, the likelihood function can be approximated by

\begin{equation} 
    P(\hat {x}_{k}|x)\simeq \frac {1}{\pi \eta _{k}^{2}}\text {exp}\left ({-\frac {1}{\eta _{k}^{2}}\left \vert{ \hat {x}_{k} -\mu _{k}}\right \vert ^{2}}\right). 
\end{equation}

\subsection{Overall Algorithm}
The pseudo-code of the proposed RIS-assisted IDD scheme is described in
Algorithm 1.
\vspace{-0.5em}
\begin{algorithm}[H]
    \label{algor1}
    \begin{algorithmic}[1]
        \caption{Proposed RIS + IDD Scheme}\label{alg:cap}
        \STATE \textbf{Input:} Channels: $\mathbf{G, F, H}$ and received signal: $\mathbf{y}$.
        \STATE \textbf{Output:} Optimized linear detection filter $\mathbf{W}$, optimized RIS precoding matrix $\mathbf{\Phi}$.
        \STATE Randomly initialize $\mathbf{\Phi}$.
        \STATE Update the value of $\mathbf{W}$.

        \FOR{$i=1$ to idd.iterations}
                \IF{$i=1$ (first iteration)}
                    \WHILE{No convergence or timeout}
                        \STATE Update the value of $\mathbf{\Phi}$.
                        \STATE Update the value of $\mathbf{W}$.
                    \ENDWHILE
                \ELSIF{$i> 1$}
                        \STATE \textbf{Detection Scheme - SIC}
                        \STATE {Obtain the receiver filter ($W$).}
                        \STATE {Obtain the Extrisic bit LLRs ($L_D$).}
                \ENDIF
                \STATE \textbf{Proceed with LDPC decoding.}
                \STATE Compute the LLR value $L_C$. 
        \ENDFOR
    \end{algorithmic}
\end{algorithm}

\section{Numerical Results}

In this section, we evaluate the BER and the sum-rate performances produced by the proposed algorithm. To assess the effectiveness of the proposed technique, we consider the following schemes:

\begin{itemize}
    \item \textbf{Standard MMSE MIMO:} This is a typical MIMO system without the presence of RIS and the IDD scheme (uncoded system). Here, we utilize MMSE to design the detector.
    \item \textbf{RIS-assisted MIMO:} This is a typical RIS-assisted MIMO system without the IDD scheme (uncoded system). Here, we utilize the algorithm presented in Section \ref{sec:designRIS} to optimize the phases of the RIS.
    \item \textbf{IDD MMSE MIMO:} This is an LDPC-coded system using soft MIMO detectors, without the presence of the RIS. The algorithm used was proposed in \cite{1494998}.
    \item \textbf{RIS-assisted IDD MIMO:} This is the proposed method with the presence of RIS and the IDD scheme. Algorithm in \ref{algor1} is used for LDPC decoding and the design of reflective parameters.
\end{itemize}

We considered a short length regular LDPC code \cite{memd,vfap} with block length $n=512$ and rate $R = 1/2$ with QPSK modulation. The channel is assumed to experience block fading under perfect
channel state information by the receiver.

\begin{figure}[htbp]
    \centerline{\includegraphics[width=0.33\textwidth]{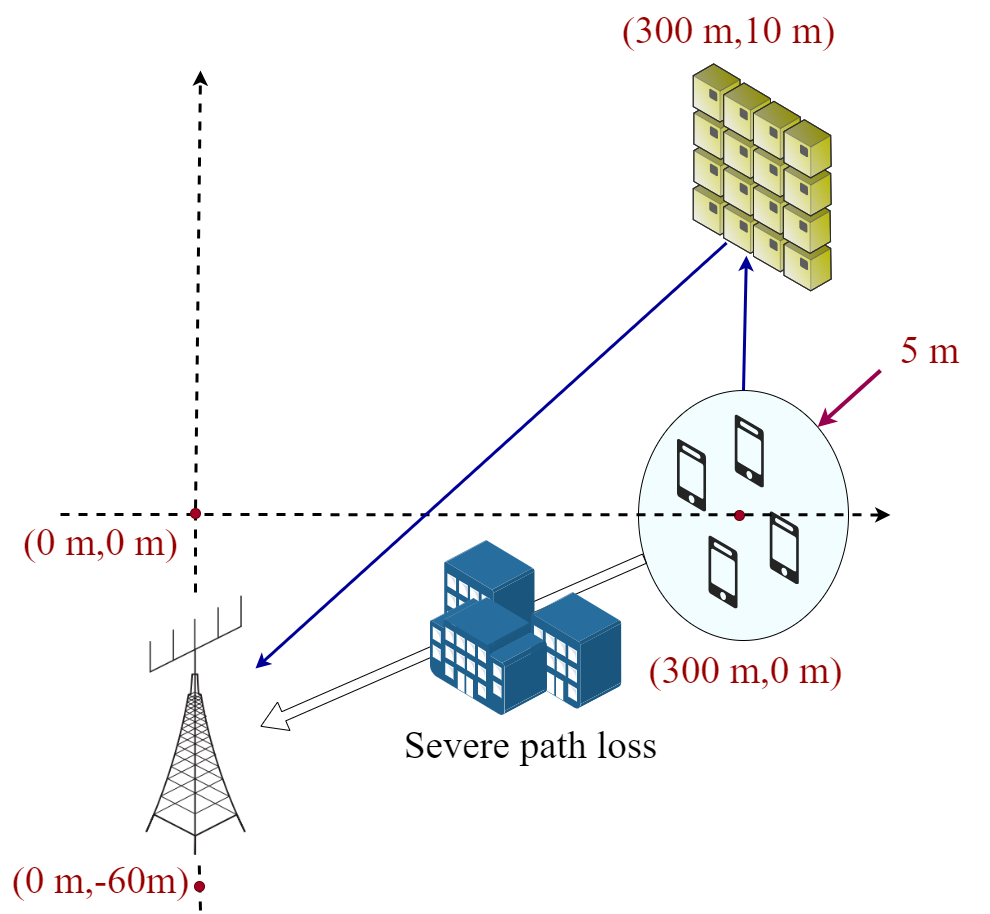}}
    \caption{Simulation scenario with weak direct link.}
    \label{dia3}
\end{figure}

It was also considered that the systems operates at a frequency of 5 Ghz, and the  direct link is weak due to severe obstruction, as illustrated in Fig \ref{dia3}. To characterize the large-scale fading of the channels the path models loss are from the the 3GPP standard \cite{access2010further}.
\begin{align} 
{\mathrm{P}}{{\mathrm{L}}_{w}}=&41.2 + 28.7\log d,
\\
{\mathrm{P}}{{\mathrm{L}}_{s}}=&37.3 + 22.0\log d.
\end{align}

The path loss model $\text{PL}_w$ is employed to simulate the weak AP-user connection, whereas $\text{PL}_s$ is applied to model the strong connections between AP-RIS and users-RIS channels. To incorporate the effects of small-scale fading, the Rayleigh fading channel model is adopted for all the channels under consideration.
\vspace{-0.5cm}
\vspace{-0.15cm}
The selected position of the AP and the RIS is at coordinates (0, -60 m) and (300 m, 10 m), respectively. User locations are randomly distributed within a circle with a 5 m radius centered at (L, 0). The scenario involves a total of $K=12$ users, with $M=32$ AP antennas and $N=64$ RIS elements. The noise power is specified as $\sigma^2 =$-100 dBm , and the transmit power per user is normalized by the code rate, denoted as $P_u = P_{tx}R$.

The positions setup was determined by the parameters present in the downlink example from \cite{9998527}. The specific choice of the horizontal distance from the users to the origin was made because this position demonstrated the highest sum-rate performance, as depicted in \cite{9998527}.

\begin{figure}[htbp]
    \vspace{-1.5em}
    \centerline{\includegraphics[width=0.5\textwidth]{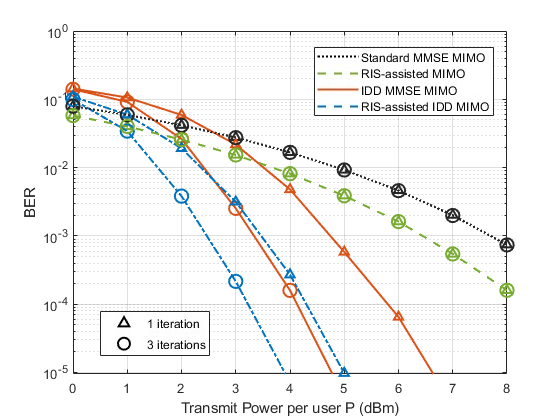}}
    \vspace{-0.5em}
    \caption{$K = 12$ and $M = 32$. BER performances of IDD schemes under the
perfect CSI.}
    \label{berfig}
    \vspace{-1.5em}
\end{figure}

The BER performances is illustrated in Fig. \ref{berfig}. It is evident that both IDD schemes achieve a substantial improvement as compared to the Standard MMSE-MIMO, particularly when employing an IDD scheme with a higher number of iterations. Moreover, the performance observed for the proposed RIS-assisted IDD-MIMO technique overcome the peformance of the IDD-MMSE-MIMO, particularly in scenarios where there is a significant loss in the direct line-of-sight path. The proposed scheme with three iterations outperforms the IDD-MMSE-MIMO scheme by approximately 2.7 dB in terms of transmit power per user, while maintaining the same BER performance.

\begin{figure}
[htbp]
    \vspace{-0.05em}
    \centerline{\includegraphics[width=0.5\textwidth]{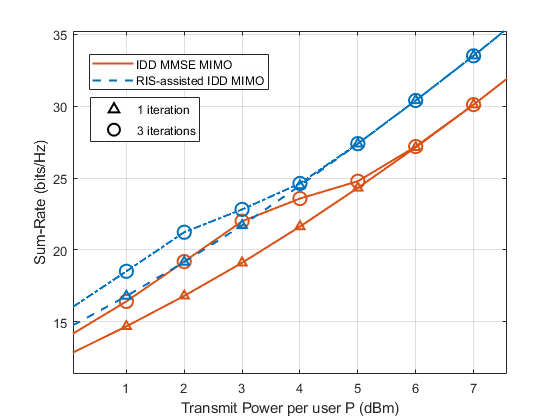}}
    \vspace{-0.05em}
    \caption{$K = 12$ and $M = 32$. Sum-rate of IDD schemes at the output of the detector under the perfect CSI.}
    \vspace{-0.1em}
    \label{capacityfig}
\end{figure}

In Fig. \ref{capacityfig}, we illustrate the sum-rate of users plotted against the transmit power per user at the output of the detector when aided by extrinsic information from the decoder. When comparing both schemes, it is noteworthy that we observe performance improvement due to the presence of the RIS, achieving a reduction of up to 1 dB in transmit power for the same sum rates. 

Additionally, it is worth noting that variations in the number of iterations within the IDD algorithm predominantly impact performance in the lower transmit power range. Beyond transmit power levels of 4 dBm and 5 dBm per user, the proposed RIS-assisted IDD-MIMO and IDD MMSE-MIMO, respectively, exhibit comparable performance for different iteration counts. Thus, we can assume that the additional improvement for higher transmit power values comes from the decoder aided by the extrinsic information $L_D$.
    \vspace{-0.075em}

\section{Conclusion}

In this paper, we introduced an IDD scheme for an RIS-assisted MIMO system in block-fading channels. The simulation results have demonstrated that the proposed algorithm leads to a significant BER performance gain after only three iterations. Additionally, a capacity analysis was conducted, comparing different methods and highlighting the attractiveness of using IDD in uplink systems, achieving up to a 1 dB reduction in transmit power per user for the same capacity when  compared with IDD MMSE MIMO. The efficacy of the proposed RIS-IDD algorithm is particularly evident in scenarios characterized by severe Line-of-Sight (LOS) path loss, providing valuable insights for optimizing the performance of RIS-assisted MIMO systems.

%\bibliographystyle{IEEEtran}
%\bibliography{refs}

\begin{thebibliography}{100} 
 
\bibitem{mmimo}
R. C. de Lamare. “Massive MIMO systems: signal processing
challenges and future trends”. In: URSI Radio
Science Bulletin 347 (2013), pp. 8–20.

\bibitem{wence} 
W. Zhang et al. “Large-Scale Antenna Systems With
UL/DL Hardware Mismatch: Achievable Rates Analysis
and Calibration”, IEEE Transactions on Communications
63.4 (2015), pp. 1216–1229. 

\bibitem{rista}
Y. Zhao and J. He. RISTA - Reconfigurable Intelligent
Surface Technology White Paper (2023). Mar. 2023.

\bibitem{9140329} 
M. Di Renzo et al. “Smart Radio Environments Empowered
by Reconfigurable Intelligent Surfaces: How
It Works, State of Research, and The Road Ahead”, IEEE Journal on Selected Areas in Communications
38.11 (2020), pp. 2450–2525.

\bibitem{9359653}
J. Ye, A. Kammoun, and M. -S. Alouini. “Spatially-
Distributed RISs vs Relay-Assisted Systems: A Fair
Comparison”. IEEE Open Journal of the Communications
Society 2 (2021), pp. 799–817.

\bibitem{9548940} 
M. Forouzanmehr, A. Khalili S. Akhlaghi, and Q. Wu.
“Energy Efficiency Maximization for IRS-Assisted Uplink
Systems: Joint Resource Allocation and Beamforming
Design”, IEEE Communications Letters 25.12
(2021), pp. 3932–3936.

\bibitem{9998527} Z. Zhang et al. “Active RIS vs. Passive RIS: Which Will Prevail in 6G?”, IEEE Transactions on Communications 71.3 (2023), pp. 1707–1725.

\bibitem{9913356} 
H. Li, S. Shen, and B. Clerckx. “Beyond Diagonal
Reconfigurable Intelligent Surfaces: From Transmitting
and Reflecting Modes to Single-, Group-, and Fully-
Connected Architectures”, IEEE Transactions on
Wireless Communications 22.4 (2023), pp. 2311–2324.

\bibitem{9452133} 
Y. Hu, Z. Lin P. Wang, and M. Ding. “Performance
Analysis of Reconfigurable Intelligent Surface Assisted
Wireless System With Low-Density Parity-Check
Code”. IEEE Communications Letters 25.9 (2021),
pp. 2879–2883.

\bibitem{9110912}
S. Zhang and R. Zhang. “Capacity Characterization for Intelligent Reflecting Surface Aided MIMO Communication”, IEEE Journal on Selected Areas in Communications 38.8 (2020), pp. 1823–1838.

\bibitem{1494998} 
A. Matache, C. Jones, and R. Wesel. “Reduced complexity
MIMO detectors for LDPC coded systems”, IEEE MILCOM 2004. Military Communications Conference, 2004. Vol. 2. 2004, 1073–1079 Vol. 2. 

\bibitem{8240730}
Z. Shao, R. C. de Lamare, and L. T. N. Landau. “Iterative
Detection and Decoding for Large-Scale Multiple-
Antenna Systems With 1-Bit ADCs”, IEEE Wireless
Communications Letters 7.3 (2018), pp. 476–479.

\bibitem{7105928} 
A. G. D. Uchoa, C. T. Healy, and R. C. de Lamare.
“Iterative Detection and Decoding Algorithms for
MIMO Systems in Block-Fading Channels Using LDPC
Codes”, IEEE Transactions on Vehicular Technology
65.4 (2016), pp. 2735–2741.

\bibitem{idd_ris1} R. C. G. Porto and R. C. de Lamare. “Iterative Detection and Decoding for RIS-Assisted Multiuser Multiple-Antenna Systems”, 19th International Symposium on Wireless Communication Systems (ISWCS).
2024, pp. 1–5. DOI: 10 . 1109 / ISWCS61526 . 2024 .
10639075.

\bibitem{idd_ris2}
R. C. G. Porto and R. C. de Lamare. “Iterative Detection
and Decoding for Multiuser Systems Based on MMSE
Refinements with Active or Passive RIS”, IEEE
Wireless Communications Letters (2024), pp. 1–5.

\bibitem{access2010further}
Evolved Universal Terrestrial Radio Access. “Further
advancements for E-UTRA physical layer aspects (Release
9)”. In: European Telecommunications Standards
Institute (2010).

\bibitem{774855} 
X. Wang and H. V. Poor. “Iterative (turbo) soft interference
cancellation and decoding for coded CDMA”, 
IEEE Transactions on Communications, 1999.

\bibitem{tds1} 
P. Clarke and R. C. de Lamare. “Joint Transmit Diversity Optimization and Relay Selection for
Multi-Relay Cooperative MIMO Systems Using Discrete
Stochastic Algorithms”, IEEE Communications
Letters, vol. 15, no. 10, 2011, pp. 1035–1037. 

\bibitem{tds2}
P. Clarke and R. C. de Lamare, 
“Transmit Diversity and Relay Selection Algorithms for Multirelay Cooperative MIMO Systems”, IEEE Transactions on
Vehicular Technology, vol. 61, no. 3, 2012, pp. 1084–1098. 

\bibitem{jidf}
R. C. de Lamare and R. Sampaio-Neto, “Adaptive
Reduced-Rank Processing Based on Joint and Iterative
Interpolation, Decimation, and Filtering”, 
IEEE Transactions on Signal Processing, vol. 57, no. 7, 2009, pp. 2503–2514. 

\bibitem{spa} 
R. C. De Lamare and R. Sampaio-Neto. “Minimum
Mean-Squared Error Iterative Successive Parallel Arbitrated
Decision Feedback Detectors for DS-CDMA
Systems”, IEEE Transactions on Communications, vol. 56, no. 5, 2008, pp. 778–789. 

\bibitem{mfsic}
P. Li, R. C. de Lamare, and R. Fa. “Multiple Feedback
Successive Interference Cancellation Detection for
Multiuser MIMO Systems”, IEEE Transactions on
Wireless Communications, vol. 10, no. 8, 2011, pp. 2434–2439.

\bibitem{mbdf}
R. C. de Lamare, “Adaptive and Iterative Multi-Branch
MMSE Decision Feedback Detection Algorithms for
Multi-Antenna Systems”, IEEE Transactions on
Wireless Communications, vol. 12, no. 10, 2013, pp. 5294–5308.

\bibitem{dfcc}
Peng Li and Rodrigo C. De Lamare, 
“Adaptive Decision-Feedback Detection With Constellation Constraints for MIMO Systems”, IEEE Transactions on
Vehicular Technology, vol. 61, no. 2,  2012, pp. 853–859. 

\bibitem{did}
P. Li and R. C. de Lamare. “Distributed Iterative Detection
With Reduced Message Passing for Networked MIMO Cellular Systems”, IEEE Transactions on Vehicular Technology, vol. 63, no. 6, 2014, pp. 2947–2954. 

\bibitem{1bitidd}
Z. Shao, R. C. de Lamare, and L. T. N. Landau, “Iterative Detection and Decoding for Large-Scale Multiple-Antenna Systems With 1-Bit ADCs”, IEEE Wireless Communications Letters, vol. 7, no. 3, 2018, pp. 476–479.

\bibitem{listmtc} 
R. B. Di Renna and R. C. de Lamare, “Iterative List
Detection and Decoding for Massive Machine-Type
Communications”, IEEE Transactions on Communications
vol. 68, no. 10, 2020, pp. 6276–6288. 

\bibitem{detmtc}
Roberto B. Di Renna et al. “Detection Techniques for
Massive Machine-Type Communications: Challenges
and Solutions”, IEEE Access, vol. 8, 2020, pp. 180928–
180954. 

\bibitem{msgamp1} 
R. B. Di Renna and R. C. de Lamare, “Dynamic
Message Scheduling Based on Activity-Aware Residual
Belief Propagation for Asynchronous mMTC”, 
IEEE Wireless Communications Letters, vol. 10, no. 6, 2021,
pp. 1290–1294. 

\bibitem{msgamp2}
R. B. Di Renna and R. C. de Lamare, “Joint Channel
Estimation, Activity Detection and Data Decoding
Based on Dynamic Message-Scheduling Strategies for
mMTC”, IEEE Transactions on Communications, vol. 
70, no. 4, 2022, pp. 2464–2479. 

\bibitem{comp}
A. B. L. B. Fernandes et al, “Multiuser-MIMO Systems
Using Comparator Network-Aided Receivers With 1-Bit
Quantization”, IEEE Transactions on Communications
vol. 71, no. 2, 2023, pp. 908–922. 

\bibitem{memd}
C. T. Healy and R. C. de Lamare, “Design of LDPC
Codes Based on Multipath EMD Strategies for Progressive
Edge Growth”, IEEE Transactions on Communications
vol. 64, no. 8, 2016, pp. 3208–3219. 

\bibitem{vfap} 
J. Liu and R. C. de Lamare, “Low-Latency Reweighted
Belief Propagation Decoding for LDPC Codes”, 
IEEE Communications Letters, vol. 16, no. 10, 2012, pp. 1660–1663.

\end{thebibliography}
%\printbibliography

\end{document}